\documentclass[iop]{emulateapj}
\usepackage{color}

\begin{document}

\title{{On the Short GRB GeV emission from a Kerr Black hole}}

\author{R.~Ruffini\altaffilmark{1,2,3,4},
M.~Muccino\altaffilmark{5},
Y.~Aimuratov\altaffilmark{1,2,6},
M.~Amiri\altaffilmark{3,8},
C.~L.~Bianco\altaffilmark{1,3},
Y.-C.~Chen\altaffilmark{1,3},
B.~Eslam Panah\altaffilmark{3,9}
G.~J.~Mathews\altaffilmark{7},
R.~Moradi\altaffilmark{1,3},
G.~B.~Pisani\altaffilmark{1,3},
D.~Primorac\altaffilmark{1,3},
J.~A.~Rueda\altaffilmark{1,3,4},
Y.~Wang\altaffilmark{1,3}
}

\altaffiltext{1}{ICRA and Dipartimento di Fisica, Universit\`a  di Roma ``La Sapienza'', Piazzale Aldo Moro 5, I-00185 Roma, Italy}
\altaffiltext{2}{Universit\'e de Nice Sophia-Antipolis, Grand Ch\^ateau Parc Valrose, Nice, CEDEX 2, France} 
\altaffiltext{3}{International Center for Relativistic Astrophysics Network, Piazza della Repubblica 10, I-65122 Pescara, Italy}
\altaffiltext{4}{ICRANet-Rio, Centro Brasileiro de Pesquisas F\'isicas, Rua Dr. Xavier Sigaud 150, 22290--180 Rio de Janeiro, Brazil}
\altaffiltext{5}{Istituto Nazionale di Fisica Nucleare, Laboratori Nazionali di Frascati, I-00044 Frascati, Italy}
\altaffiltext{6}{Fesenkov Astrophysical Institute, Observatory 23, 050020 Almaty, Kazakhstan}
\altaffiltext{7}{Center for Astrophysics, Department of Physics, University of Notre Dame, Notre Dame, IN, 46556, USA}

\altaffiltext{8}{Department of Physics, Isfahan University of Technology, 84156-83111, Iran}
\altaffiltext{9}{Physics Department and Biruni Observatory, College of Sciences, Shiraz University, Shiraz 71454, Iran}

\begin{abstract}
{It has recently become clear that in both short and long gamma-ray bursts (GRBs) it coexists a sequence of different events, each characterized by specific physical processes and corresponding values of the Lorentz gamma factors. The ultra-relativistic prompt emission (UPE) phase, with Lorentz factor $\Gamma\leq10^4$, is followed by a mildly relativistic plateau-afterglow phase with $\Gamma\lesssim2$. The GeV radiation, with $\Gamma\lesssim50$, coexists with the above two phases. It is shown that: a) the GeV radiation originates at the onset of the formation of a black hole (BH), b) its luminosity follows specific power-law dependence when measured in the rest frame of the source with a decay index $\gamma=-1.29\pm0.06$ in the case of the short GRBs, and $\gamma=-1.20\pm0.04$ in the case of the long GRBs, c) these energetics requirements are used to derive the mass and spin of the BH originating this extended GeV emission. We present these conceptual results for the case of short GRBs in this article and give the extended analysis for long GRBs in a companion article. A direct astrophysical application of these results is that the merger of binary neutron stars leading to BH formation emits GeV radiation: the GeV emission is a necessary and sufficient condition to indicate the creation of a BH in S-GRBs.}
\end{abstract}


\section{Introduction}

Following the discovery of gamma-ray bursts (GRBs) by the Vela satellite \citep{1973ApJ...182L..85K}, the Compton Gamma-Ray Observatory (CGRO) provided the first spectral and temporal burst dataset which led {to their} classification into short and long bursts (\citealt{1981Ap&SS..80....3M,Klebesadel1992,Dezalay1992,Koveliotou1993,Tavani1998}, see also \citealt{1981Ap&SS..80....3M}).

With the {1996 launch of the Beppo-Sax satellite}, X-rays were added to the prompt emission gamma-ray observations, leading to three {fundamental new} discoveries that can be used as GRB diagnostics: 
1) the X-ray afterglow \citep{Costa1997}, which allowed more precise GRB localizations in the optical bands \citep{vanParadjis1997}, 
2) the determination of their cosmological redshifts \citep{1997Natur.387..878M}, which led to the proof of their cosmological nature and 
3) the spatial and temporal coincidence of some long bursts with supernovae \citep[SNe, see][]{1998Natur.395..670G}. 

These ground-breaking discoveries were soon enhanced by the observations of the \textit{Swift} satellite \citep{2004ApJ...611.1005G} and further extended from {the} MeV to GeV energy range by the \textit{Fermi} satellite \citep{2009ApJ...697.1071A}.

In parallel {with} this sequence of fundamental observational discoveries, the theoretical comprehension of long GRBs has made significant progress. In particular, from the earliest recognition of the expansion of the $e^+e^-$ optically thick plasma as the acceleration process {for} GRBs, the requirement of a proper general relativistic treatment led to the introduction of four different time coordinates. Furthermore, the consideration of {a} binary system was introduced to explain the GRB-SN coincidence \citep{Ruffini2001a,Ruffini2001b, Ruffini2001c}. Identification of the long GRB progenitors with {an} initial state (\textit{in-state}) tight binary system composed of a carbon-oxygen core (CO$_{\rm core}$) led to the introduction of two long GRB subclasses. The core undergoes a SN explosion forming a newborn neutron star ($\nu$NS) and hypercritically accretes onto a companion NS. From the two subclasses the first is the X-ray flashes (XRFs), with a final-state (\textit{out-state}) consisting of a $\nu$NS-NS binary, and the second, the binary-driven hypernovae (BdHNe), with a out-state comprised of a $\nu$NS-BH binary. This progress has been summarized in three articles and references therein \citep{2016ApJ...832..136R,2016ApJ...833..107B,2018ApJ...852...53R}. The fundamental role of neutrinos has been recently discussed in \citet{2018ApJ...852..120B}.

The aim of this article is to use the information gained from \textit{Swift}, \textit{Fermi} and especially \textit{Fermi}-LAT in order to probe the nature of two subclasses of short bursts: the short gamma-ray flashes (S-GRFs) and the genuine short gamma-ray bursts (S-GRBs).

The progenitor systems of short bursts have been {identified with binary NS mergers in classic papers} \citep[see, e.g.,][]{1986ApJ...308L..47G,1986ApJ...308L..43P,Eichler:1989jb,1991ApJ...379L..17N,1992ApJ...395L..83N,1538-4357-482-1-L29,2014ARA&A..52...43B}, localized at large off-sets from their host galaxies and with no star formation evidence \citep[see, e.g.,][]{Fox2005,Gehrels2005,2014ARA&A..52...43B}. This identification further evolved {with} the introduction of the two sub-classes of short bursts \citep{2015ApJ...798...10R,2016ApJ...831..178R,2016ApJ...832..136R,2017ApJ...844...83A}: A first sub-class corresponds to  short bursts with isotropic energies $E_{\rm iso}<10^{52}$~erg (in the rest-frame $1$--$10^4$~keV energy band) and rest-frame spectral peak energies $E_{\rm p,i}<2$~MeV, expected to originate when the NS-NS merger leads to a single massive NS (M-NS) with the mass below the NS critical mass. We have {called} these sources short gamma-ray flashes (S-GRFs).

The second sub-class corresponds to the short bursts with $E_{\rm iso}\gtrsim10^{52}$~erg and $E_{\rm p,i}\gtrsim2$~MeV. It is expected to originate from a NS-NS merger in which the merged core overcomes the NS critical mass and gravitationally collapses to form a BH. We have {called these} genuine short GRBs (S-GRBs). 
  
Indeed, the beginning of the prompt phase of S-GRBs exhibits GeV emission heralding the formation of the BH \citep[see, e.g.,][]{2017ApJ...844...83A,2016ApJ...831..178R}. This signature is missing in S-GRFs in which no BH is formed. For S-GRBs we derive the $0.1$--$100$~GeV luminosity light-curves as a function of time in the rest-frame of the source, and show how such {a} luminosity follows a universal power-law $L(t)= (0.88\pm 0.13) \times 10^{52}~t^{-1.29 \pm 0.06}$~erg~s$^{-1}$.

As summarized in section~\ref{sec:2}, seven different subclasses of GRBs have been identified. We update their number while recalling {the \textit{in-states} of their progenitors as well as the \textit{out-states}} in their cosmic matrix form \citep{2016arXiv160203545R}.

Properties of 18 S-GRFs are discussed in section~\ref{sec:3}. All of them {have been} observed since the launch of \textit{Fermi} {through} the end of 2016 and have {known} cosmological redshifts. We confirm that their soft {energy} spectrum {is} in the range $E_{\rm p,i}\sim0.2$--$2$ MeV and that their isotropic energies are lower than $E_{\rm iso}<10^{52}$~erg \citep{2016ApJ...832..136R}. We also confirm that {as expected} no GeV emission is observed in any of them.

In section~\ref{sec:4} we first give the prompt and GeV emission properties for each of the 6 S-GRBs observed after the launch of \textit{Fermi}, with known or derived cosmological redshifts. We confirm their harder spectra with a peak energy at $E_{\rm p,i}\sim 2$--$8$ MeV, and that their effective isotropic energy $E_{\rm iso}$ is larger than $10^{52}$~erg, as expected \citep{2016ApJ...832..136R}. For each source we give the reference in which their analysis has been done. We also present {their} $E_{\rm LAT}$ deducible from the $Fermi$-LAT light curve. Finally, we derive the universal power-law behavior of their GeV emission luminosity  as a function of  time in the rest-frame of the source, with an amplitude of $(0.88\pm 0.13)\times 10^{52}$ and a power-law index of $1.29 \pm 0.06$.  

In section~\ref{sec:5} we first recall the accretion of co-rotating or counter-rotating matter onto a Kerr BH  as the possible source of the GeV emission \citep{2016ApJ...831..178R, 2017ApJ...844...83A}. 
That approach, although energetically satisfactory, does not offer the possibility to infer either the mass or the spin of the BH. In this article, we indicate the basic equations relating the GeV emission to the spin and mass of the newly-formed BH for selected NS equations of state (EoS). 

We summarize our main results in section~\ref{sec:6},  leading to the minimum mass and corresponding maximum dimensionless spin parameter $\alpha $ for all 6 S-GRBs in the narrow ranges of $2.24$--$2.89 M_\odot$ and $0.18$--$0.33$, respectively. 
We conclude that the observation of the GeV emission gives the necessary and sufficient condition {for identifying} the existence of a BH in S-GRBs.

For the benefit of the reader, Table~\ref{acronyms} summarizes {in alphabetical order the acronyms used in this article}.

\begin{table}
\centering
\begin{tabular}{lc}
\hline\hline
Extended wording & Acronym \\
\hline
Binary-driven hypernova & BdHN \\
Binary neutron star & B-NS\\
Black hole                    & BH \\
Carbon-oxygen core      & CO$_{\rm core}$ \\ 
Gamma-ray burst         & GRB \\
Gamma-ray flash          & GRF \\
Massive neutron star     & M-NS \\
Neutron star                & NS \\
New neutron star          & $\nu$NS \\
Short gamma-ray burst  & S-GRB \\
Short gamma-ray flash  & S-GRF \\
Supernova                  & SN \\
Ultrashort gamma-ray burst & U-GRB \\ 
White dwarf                & WD \\
X-ray flash                  & XRF \\
\hline
\end{tabular}
\caption{List of acronyms used in this work.}
\label{acronyms}
\end{table}

\section{Summary on seven subclasses of the fireshell model}\label{sec:2}

In this paper we address the specific role of the GeV emission in order to further characterize the 7 subclasses of GRBs presented in \citet{2016ApJ...832..136R}. Toward this goal in the following we only consider sources observed after the launch of \textit{Fermi}, when LAT data became available. Correspondingly, we have updated the number presented in Fig.~38 of \citet{2018ApJ...852...53R}. In Table~\ref{tab:a} we have, for each of the 7 subclasses of GRBs, indicated the name, the number of observed sources with definite cosmological redshift and the progenitors characterizing the initial state. In all cases the progenitors are binary systems composed of various combinations of a CO$_{\rm core}$ undergoing a SN explosion, a $\nu$NS  created in such an explosion and other compact objects, such as a white dwarf (WD), a NS or a BH. The outcome of the {merger} process is represented in Fig.~7 in \citep{2016ApJ...832..136R}.

\begin{table*}
 \centering
   \begin{tabular}{lcccccccc}
   \hline
     & Sub-class  & Number & \emph{In-state}  & \emph{Out-state} & $E_{\rm p,i}$ &  $E_{\rm iso}$  &  $E_{\rm iso,Gev}$  \\
     & & & & & (MeV) & (erg) &  (erg)\\  		
   \hline
   I   & S-GRFs & $18$ &NS-NS & MNS & $\sim0.2$--$2$ &  $\sim 10^{49}$--$10^{52}$  &  $-$ \\
   II    & S-GRBs  & $6$ &NS-NS & BH & $\sim2$--$8$ &  $\sim 10^{52}$--$10^{53}$ &   $\gtrsim 10^{52}$\\
   III    & XRFs & $48$ &CO$_{\rm core}$-NS    & $\nu$NS-NS & $\sim 0.004$--$0.2$  &  $\sim 10^{48}$--$10^{52}$ &    $-$ \\
   IV   & BdHNe  & $329$ &CO$_{\rm core}$-NS  & $\nu$NS-BH & $\sim0.2$--$2$ &  $\sim 10^{52}$--$10^{54}$ &    $\gtrsim 10^{52}$ \\
   V  & BH-SN & $4$ &CO$_{\rm core}$-BH  & $\nu$NS-BH & $\gtrsim2$ &  $>10^{54}$ &   $\gtrsim 10^{53}$   \\
   VI   & U-GRBs & $0$ &$\nu$NS-BH & BH & $\gtrsim2$ &  $>10^{52}$ & $-$ \\
   VII  & GRFs  & $1$ &NS-WD & MNS & $\sim0.2$--$2$ &  $\sim 10^{51}$--$10^{52}$  & $-$\\
   \hline
\end{tabular}
\caption{Summary of the GRB subclasses from \citep{2016ApJ...832..136R}. In addition to the subclass name, we list the number of GRBs for each subclass updated {through} the end of 2016. We also summarize the ``in-state'' representing the progenitors and the ``out-state'', as well as the  $ E_{\rm p,i}$ and $E_{\rm iso}$  for each subclass. We indicate the GeV emission in the last column, which is  uniquely for  the BdHNe and BH-SN (in the case of  long GRBs) and for the S-GRBs (in the case of short GRBs). In all of these events the GeV emission is greater than $10^{52}$erg.}
\label{tab:a}
\end{table*}

For the S-GRFs we have updated our classification and subdivided the 18 S-GRFs into three different groups: the ones without associated \textit{Fermi} observations, the ones outside the boresight angle in the \textit{Fermi}-LAT and the ones within the boresight angle of the \textit{Fermi}-LAT; see Table~\ref{tab:ba}. Their in-states are represented by a NS-NS binary and their out-state are represented by a M-NS. The absolute lower limit for the total energy of S-GRFs is $ \sim 10^{49}$~erg. Their spectral peak energy has the range of $0.2$ MeV~$\lesssim E_{\rm p,i}\lesssim 2$~MeV and isotropic energy $10^{49}\lesssim E_{\rm iso}\lesssim 10^{52}$~erg. GeV emission is not observed in any of them.

\begin{table*}
\centering
\begin{tabular}{clccclccl}
\hline\hline
Group 	          & S-GRF	& $z$       &	$E_{\rm p}$	& $E_{\rm iso}$     & Fermi GCN & $\theta$  & GeV observed  & Comments \\
	  	          &         &           &	keV			& ($10^{50}$~erg)   &               & (deg)     &           & \\
\hline	          
	  	          & 090426	&	$2.609$	&			&	$44.5\pm6.6$	&	 $-$    	&	$-$	    &	no	    &  \\
	  	          & 090515	&	$0.403$	&			&	$0.094\pm0.014$	&	 $-$    	&	$-$	    &	no	    &  \\
	  	          & 100724A &	$1.288$	&			&	$16.4\pm2.4$	&	 $-$    	&	$-$	    &	no	    &  \\
No \textit{Fermi} & 101219A &	$0.718$	&			&	$48.8\pm6.8$	&	 $-$    	&	$-$	    &	no	    &  \\
Observation 	  & 120804A &	$1.3$	&			&	$70.0\pm15.0$	&	 $-$    	&	$-$	    &	no	    & $46^{day}$ x-ray (GCN 13841) \\
	  			  & 130603B &	$0.356$	&			&	$21.2\pm2.3$	&	 $-$    	&	$-$	    &	no	    & \textbf{kilonova} (GCN 14893, 14895, 14913) \\
	  			  & 140622A &	$0.959$	&			&	$0.70\pm0.13$	&	 $-$    	&	$-$	    &	no	    &  \\
	  			  & 140903A &	$0.351$	&			&	$1.41\pm0.11$	&	 $-$     	&	$-$	    &	no	    &  \\
\hline
	  			  & 090927	&	$1.37$	&	408.64	&	$7.6\pm3.5$	    &	GCN 9974	&	$85.0$	&	no	    & \\
Outside 		  & 100117A &	$0.915$	&	625.73	&	$78.0\pm10.0$	&	GCN 10345	&	$86.0$	&	no	    & \\
Boresight 		  & 100625A &	$0.453$	&	701.81	&	$7.50\pm0.30$	&	GCN 10912	&	$125.0$	&	no	    & \\
Angle 			  & 131004A &	$0.717$	&	202.45	&	$12.7\pm0.9$	&	GCN 15315	&	$93.0$	&	no	    & \\
				  & 141004A &	$0.573$	&	268.99	&	$21.0\pm1.9$	&	GCN 16900	&	$100.4$	&	no	    & \\
\hline
				  & 080905A	&	$0.122$	&	443.12	&	$6.58\pm0.96$	&	GCN 8204	&	$28.0$	&	no	    & \\
Inside 			  & 100206A &	$0.408$	&	748.34	&	$4.67\pm0.61$	&	GCN 10381	&	$44.7$	&	no	    & \\
Boresight 	      & 111117A &	$1.31$	&	857.40	&	$34.0\pm13.0$	&	GCN 12573	&	$12.0$	&	no	    & \\
Angle 			  & 150101B &   $0.134$ &   141.88	&	$0.4$           &   GCN 17276   &   $44.0$  &   no      & \cite{2015ApJ...815..102F}, $39^{day}$ x-ray (GCN 17431) \\
				  & 160821B &   $0.16$  &   97.44	&	$1.2$           &   GCN 19843   &   $61.0$  &   no      & \textbf{kilonova} \\
\hline
\end{tabular}
\caption{
\textit{\textbf{List of $18$ short gamma-ray flashes (S-GRFs)}} 
divided {into} three different groups: 
1) {those without a} \textit{Fermi}-LAT observation (upper group); 
2) {those with a} \textit{Fermi}-LAT observation but the boresight  $\theta \geq 75^{\circ}$, and 
3) those within the \textit{Fermi}-LAT boresight angle (lower group, $\theta < 75^{\circ}$). None of the S-GRFs have associated GeV emission detected. In the first column we indicate the name of the source, in the second their redshift, in the third column  the $E_{\rm p,i}$ deducible from the $Fermi$ data, in the fourth column we estimate the $E_{\rm iso}$ which is systematically lower than the $10^{52}$ erg. For the convenience  we also add  both the specific GCN of the $Fermi$ source as well as the boresight angle of the LAT observation. We also note the non-observation of the Gev emission. The last column contains comments about Kilonova emission. The symbol ``$-$'' indicates no information on the LAT boresight angle due to the lack of GBM observation.}
\label{tab:ba}
\end{table*}

For the S-GRBs we focus our analysis on sources detected after the launch of \textit{Fermi} (namely 6 sources): \textit{all} of them were observed by LAT and show GeV emission implying a very broad emission angle, i.e., almost isotropic, and all of them have energies larger than $ 10^{52}$~erg. {Their in-state correspond} to a NS-NS binary and their out-state consists of a single BH. The spectral peak energy is in the range of $2\: \rm MeV\lesssim E_{\rm p,i}\lesssim 8$~MeV, with their isotropic energy in the range of $10^{52}\lesssim E_{\rm iso}\lesssim 10^{53}$~erg, with $E_{\rm iso,GeV}>10^{51}$~erg. More details {are given below} in section~\ref{sec:4}.

After the launch of \textit{Fermi}, $48$ sources  have been {designated} as XRFs, with their ``in-state" represented by a CO$_{\rm core}$-NS binary and their out-state represented by a  $\nu$NS, originated {from} the SN explosion of a CO$_{\rm core}$ and a companion NS. Their spectral peak energy is in the range of $4~\rm keV\lesssim E_{\rm p,i}\lesssim 660$~keV and their isotropic energy in {the}  $10^{48} \lesssim E_{\rm iso}\lesssim 10^{52}$~erg range. No GeV emission has been observed in these sources. The complete list of XRFs which updates the one in \citet{2016ApJ...832..136R} is given in the accompanying paper ``On the universal GeV emission in BdHNe and their inferred morphological structure'' \citep{Ruffini2018arXiv180305476R}.

The first list of BdHNe was introduced in \citet{2016ApJ...833..159P, 2016ApJ...832..136R}, further extended in \citet{2018ApJ...852...53R}, {and} still further extended in the accompanying paper \citet{Ruffini2018arXiv180305476R}. {329} BDHNe sources have been identified after the launch of \textit{Fermi}, all with {known redshift}. Their in-states are represented by a CO$_{\rm core}$-NS binary and their out-states represented by a  $\nu$NS, originating in the SN explosion of a CO$_{\rm core}$, and a companion BH. Their spectral peak energy has the range of $0.2~\rm MeV\lesssim E_{\rm p,i}\lesssim 2$~MeV. Their isotropic energy is in the range of $10^{52}\lesssim E_{\rm iso}\lesssim 10^{54}$~erg and their isotropic GeV emission is $\sim 10^{53}$~erg. Two subclasses of BdHNe have been found corresponding to the  presence or absence of the GeV emission.

Concerning the BH-SN systems, they are a subset of BdHNe group, corresponding to particularly energetic sources and a BH with a larger mass, all the way {up} to 29~M$_{\odot}$. Their progenitor is a CO$_{\rm core}$-BH binary and their out-state consists of a  $\nu$NS, originating in the SN explosion of a CO$_{\rm core}$ and a companion BH. Their spectral peak energy is larger than 2~MeV. Their isotropic energies are $E_{\rm iso}>10^{54}$~erg and their isotropic GeV emission is $\sim 10^{53}$~erg. Details will be given in a forthcoming paper \citep{Ruffini2018arXiv180305476R}.

U-GRBs are expected to originate from the remnant of a BdHN when the $\nu $NS merges inside the newly born BH. We have not yet observed the U-GRBs \citep{2016ApJ...832..136R,2015PhRvL.115w1102F}. From a theoretical standpoint they must exist, still we have not yet been able to identify these sources, their spectral distribution nor their timescale, while in principle, they should be as common as BdHNe \citep{2016ApJ...832..136R}. Indeed, the U-GRBs are of great interest for their conceptual novelty, and since their T$_{90}$ still has to be identified, their identification represents an additional challenge.

Concerning GRFs, 13 have been identified with only $1$ after the launch of \textit{Fermi} (see Table~\ref{tab:ba2} which gives their cosmological redshift $z$, peak energy $E_{\rm p,i}$ and isotropic energy $E_{\rm iso}$). Here we have updated their number since the previous publication (Table 11 in \citealp{2016ApJ...832..136R}). Their in-state is a binary merger of a NS with a massive WD \citep{1992A&A...266..232D,1994A&A...287..403D} and their out-state is a M-NS. Spectral peak energy is in {the}  $0.2$~MeV$\lesssim E_{\rm p,i}\lesssim 2$~MeV range, and their isotropic energy in the range of $10^{51}\lesssim E_{\rm iso}\lesssim 10^{52}$~erg. The range of the cosmological redshift is $0.089\leq z\leq2.31$ (see Table~\ref{tab:ba2}). {Note that many such WD-NS systems are expected to occur in larger numbers} than the NS-NS events \citep{2016arXiv160203545R,2018.Rueda...}. Possibly their emission is below the threshold of \textit{Swift} and \textit{Fermi} and a new class of soft X-ray missions should be envisaged. For this subclass of GRBs no SN association is expected nor observed, {even} in the case of nearby sources \citep{2006Natur.444.1050D}. GRB 060614 has {proven to be the more clearcut} case \citep{2009A&A...498..501C}, therefore becoming of great relevance for the possible role of such binary systems as progenitors of kilonova events \citep{2015NatCo...6E7323Y,2018.Rueda...}. 

\begin{table*}
\centering
\begin{tabular}{clccclccl}
\hline\hline
Group 	          & GRF	    & $z$       &	$E_{\rm p}$	& $E_{\rm iso}$     & GCN and reference    \\
	  	          &         &           &	keV			& ($10^{50}$~erg)   &                     &          \\
\hline
\hline	
                  &050724   &$0.257$    &	            &$6.19\pm0.74$		&N12                  &\\
                  &050911   &$0.165$	&               &$0.89\pm0.16$	    &N12                  &\\
                  &051227	&$0.714$	&               &$51.44\pm0.29$	    &GCN 4400             &\\
                  &060505	&$0.089$    &	            &$2.35\pm0.42$	    &GCN 5142             &\\
                  &060614	&$0.125$    &	            &$21.7\pm8.7$ 	    &N12                  &\\
                  &061006	&$0.438$	&               &$17.9\pm5.6$       &N12                  &\\
pre-\textit{Fermi} era &061021	&$0.3462$   &	            &$50\pm11$	        &GCN 5744-48          &\\
                  &061210   &$0.409$	&               &$0.24\pm0.06$   	&N12                  &\\
                  &070506	&$2.31$	    &               &$51.3\pm5.4$	    &GCN 6376             &\\
                  &070714B  &$0.923$	&               &$98\pm24$	        &N12                  &\\
                  &071227	&$0.381$    &	            &$8.0\pm1.0$	    &N12                  &\\
                  &080123	&$0.495$    &	            &$11.7\pm3.9$  	    &GCN 7205             &\\
\hline
no \textit{Fermi}-GBM  &150424A	&$0.30$	&&$24.4\pm1.6$		&GCN 17752-61             &\\
\hline
\end{tabular}
\caption{Updated list of  gamma-ray flashes (GRFs) with known redshift from \citet{2016ApJ...832..136R}. There is no \textit{Fermi} observation for this subclass. N12 = Norris, J.P., et al., 2010, ApJ, 717, 411.			
}
\label{tab:ba2}
\end{table*}

In the following, the main objective is to see how the GeV radiation can be used to discriminate between S-GRFs and S-GRBs. This {will be  accomplished by developing} a basic procedure for the determination of the BH mass and spin, see section~\ref{sec:5}. 

\section{Short gamma-ray flashes (S-GRF) }\label{sec:3}

In Table~\ref{tab:ba} we report on $18$ short gamma-ray flashes (S-GRFs) with known redshifts \citep[extending the previous catalog taken from][]{2016ApJ...832..136R}.  These include sources observed after 2008, when \textit{Fermi} started to operate, {through} the end of 2016.

Of these 18 S-GRFs, 8 were not triggered by \textit{Fermi}. All of them were  observed by \textit{Swift}, and in some cases by \textit{Konus-Wind} (upper group). The most interesting case  is the kilonova 130603B which  historically is the first kilonova  ever observed \citep{2013Natur.500..547T} which has essential complementary data in the optical and infrared to analyze the B-NS {merger}. The remaining 10 sources have been observed both by \textit{Fermi} and by \textit{Swift}. However, only  5 sources are outside the \textit{Fermi}-LAT boresight angle (middle group, $\theta \geq 75^{\circ}$), and 5 are within the \textit{Fermi}-LAT boresight angle (lower group, $\theta < 75^{\circ}$). None of the last 5 exhibit associated GeV emission. In all of them the $E_{\rm iso} $ is systematically lower than $10^{52}$ ergs. In the first column of Table~\ref{tab:ba} we indicate the name of the source, in the second their redshift and in third column  the $E_{\rm p,i}$ deduced from the \textit{Fermi} data. In the fourth column we estimate $E_{\rm iso}$, which is systematically lower than $10^{52}$~erg. For convenience, we also add both the specific GCN of the \textit{Fermi} source as well as the boresight angle of the LAT observation in the column of the non-observation of the GeV emission. Where existing, we have  indicated the evidence for associated kilonovae emission,  particularly for the case of 160821B  which presents some specific kilonova features.

\section{Short gamma-ray bursts (S-GRBs)}\label{sec:4}

Here we give short summary for each of 6 S-GRBs: 081024B, 090227B, 090510A, 140402A, 140619B and 160829A. We then  extend  our previous considerations of the specific  luminosity of the GeV emission dependence as a function of  time evaluated in the rest-frame of these sources, {using a technique already employed in our previous papers} \citep{2016ApJ...831..178R, 2017ApJ...844...83A}.

Finding the P-GRB and the prompt emission of the GRBs using \textit{Fermi}-GBM is crucial to infer the source cosmological redshift and consequently, all the physical properties of the $e^+e^-$ plasma at the transparency radius \citep{2013ApJ...763..125M,2015ApJ...808..190R,2010ApJ...712..558A}. Two observational values are needed {to derive} the source redshift: the observed P-GRB temperature $kT$ and the ratio between the P-GRB fluence and the total {fluence}. For more details on the method see \citet{2013ApJ...763..125M}. 

\begin{figure}
\centering
\includegraphics[width=1\hsize,clip]{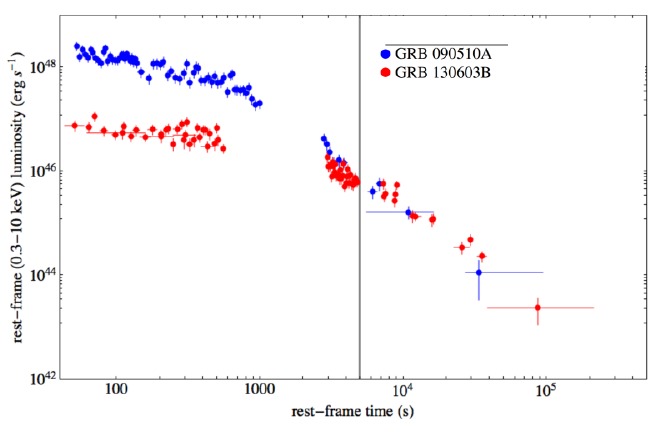}

\caption{The rest-frame $0.3$--$10$~KeV luminosity light-curves of S-GRBs 130603B and 090510. The overlap between the X-ray emissions of the two GRBs can be seen at late times.}
\label{overlap}
\end{figure}

\begin{enumerate}
\item \textit{GRB 081024B:} This source had the first clear detection of  GeV emission from a short duration gamma-ray burst \citep{2010ApJ...712..558A}.
Applying the fireshell model to this S-GRB, the {derived redshift is found} to be $z=3.12\pm1.82$ and therefore $E_{\rm iso}=(2.64\pm 1.0)\times10^{52}$~erg, $E_{\rm p,i}=(9.56\pm4.94)$~MeV, and $E_{\rm LAT}=(2.56\pm 1.5)\times10^{52}$~erg. This GRB has $E^{tot}_{e^+e^-}=2.6 \pm 1.0 \times 10^{52}$~erg and the baryon load of $B= (4.6 \pm 2.8) \times 10^{-5}$. The average circumburst medium (CBM) number density inferred from the prompt emissions of GRB 081024B is $\langle n_{\rm CBM} \rangle=(3.18\pm0.74)\times10^{-4}$~cm$^{-3}$, typical of S-GRB galactic halo environments. For more details see Table.~2 in \citet{2017ApJ...844...83A}.

\item \textit{GRB 090227B:} This is the prototype of genuine short gamma-ray bursts showing the presence of GeV emission. The theoretically deduced redshift using the fireshell model  of $z=1.61\pm 0.14$. From that, we inferred $E_{\rm iso}=(2.83 \pm 0.15) \times 10^{53}$~erg, $E_{\rm p,i}=(5.89\pm 0.30)$~MeV, and $E_{\rm LAT}=(5.78\pm0.60)\times10^{52}$~erg. This GRB has $E^{tot}_{e^+e^-}=(2.83 \pm 0.15) \times 10^{53}$~erg and {a} baryon load of $B= (4.13 \pm 0.05) \times 10^{-5}$. The CBM number density inferred from the prompt emission is $\langle n_{\rm CBM} \rangle=(1.90\pm0.20)\times10^{-4}$~cm$^{-3}$ see Table.~5 and Table.~6 in \citet{2013ApJ...763..125M}

\item \textit{GRB 090510A:} This is the brightest short gamma-ray burst with emission in the whole electromagnetic range and with a spectroscopic redshift of $z=0.903\pm 0.003$ and  a theoretically deduced redshift using  the fireshell model of $z=0.75\pm 0.17$, giving  $E_{\rm iso}=(3.95\pm0.21)\times10^{52}$~erg, $E_{\rm p,i}=(7.89\pm0.76)$~MeV and $E_{\rm LAT}=(5.78\pm0.60)\times10^{52}$~erg. This GRB has $E^{tot}_{e^+e^-}=(3.95 \pm 0.21) \times 10^{52}$~erg and {a} baryon load of $B= (5.54 \pm 0.70) \times 10^{-5} $. The CBM number density inferred from the prompt emission is $\langle n_{\rm CBM} \rangle=(8.7\pm 2.1)\times10^{-6}$~cm$^{-3}$, see Table.~3 in \citet{2016ApJ...831..178R} for more details. The overlap between the late X-ray emission of 090510A and that of 130603B, Fig.~\ref{overlap}, was the object of interest in the GCN 14913 \Citep{GCN1493...}. {The work} of \citet{2013Natur.500..547T} reported on a possible kilonova of GRB 130603B. This together with the overlapping between the optical emission of these two sources at time $\sim$ 800 s \citep{2018.Rueda...},
gives support to the possibility that {a} kilonova process may occur in the final evolution of a B-NS leading to a BH in addition to the more common one of a B-NS leading to a M-NS \citep{2013Natur.500..547T}.

\item \textit{GRB 140402A:} The theoretically derived  redshift from the fireshell model for this S-GRB is $z=5.52\pm0.93$. This implies the $E_{\rm iso}=(4.7\pm1.1)\times10^{52}$~erg, $E_{\rm p,i}=(6.1\pm1.6)$~MeV, and $E_{\rm LAT}=(4.5\pm2.2)\times10^{52}$~erg. GRB 140402A has $E^{tot}_{e^+e^-}=(4.7 \pm 1.1) \times 10^{52}$~erg and {a} baryon load is $B= 3.6 \pm 1.0 \times 10^{-5} $. The average CBM number density in the case of GRB 140402A is $\langle n_{\rm CBM} \rangle=(1.54\pm0.25)\times10^{-3}$~cm$^{-3}$.  For more details see Table.~4 in \citep{2017ApJ...844...83A}. A long-lived GeV emission within $800$~s has been reported \citep{2014GCN.16069....1B}. 

\item \textit{GRB 140619B:} The theoretically derived  redshift from the fireshell model is $z=2.67\pm0.37$. This implies: $E_{\rm iso}=(6.03\pm0.79)\times10^{52}$~erg,  $E_{\rm p,i}=(5.34\pm0.79)$~MeV and $E_{\rm LAT}=(2.34\pm0.91)\times10^{52}$~erg. GRB 140402A has $E^{tot}_{e^+e^-}=(6.03\pm0.79) \times 10^{52}$~erg and {a} baryon load is $B= (5.52 \pm 0.73) \times 10^{-5}$. The average CBM number density in the case of GRB 140619B is $\langle n_{\rm CBM} \rangle=(4.70\pm 1.20)\times10^{-3}$~cm$^{-3}$.  For more details see Table.~3 in \citet{2017ApJ...844...83A,2015ApJ...808..190R}.

\item \textit{GRB 160829A:} The theoretically inferred  redshift from the fireshell model  is $z=4.373\pm 0.99$. This implies that $E_{\rm iso}=(2.56\pm0.22)\times10^{52}$~erg, $E_{\rm p,i}=(0.92\pm 0.34)$~MeV and $E_{\rm LAT}=(4.43\pm3.86)\times10^{52}$~erg. This GRB has $E^{tot}_{e^+e^-}=(2.56\pm0.22) \times 10^{52}$~erg and {a} baryon load is $B= (1.59 \pm 0.10) \times 10^{-4}$. More details on this source will be published elsewhere.
\end{enumerate}

The values of $E_{\rm LAT}$ are calculated by multiplying the average luminosity in each time bin by the corresponding rest-frame time duration and {then summing all bins}. We must point out that since at late time the GeV emission observation could be prevented due to the instrument threshold of the LAT; these values are the lower limits to the GeV isotropic energies.

In Table~\ref{tab:bb} we summarize the prompt and the GeV emission properties for six S-GRBs, i.e., the redshift $z$, $E_{\rm p,i}$, $E_{\rm iso}$, $E_{\rm LAT}$, the position of the source from the LAT boresight $\theta$ and the likelihood test statistic (TS, to ascertain that the GeV photons belong to the burst and not to the $\gamma$-ray background sources).

\begin{table*}
\centering
\begin{tabular}{lccccccccccc}
\hline\hline
Source        	&  $z$     &  $E_{\rm p,i}$   &  $E_{\rm iso}$   & Fermi GCN &  $E_{\rm LAT}$         &  $\theta$  &  TS    & &    $comments$\\
        	&          &  (MeV)           &  ($10^{52}$~erg)  &  & ($10^{52}$~erg)       &  (deg)     &         \\ 
\hline
081024B 	&  $3.12$  &  $9.56\pm4.94$   &  $2.64\pm1.00$    & $8407$, $8408$ &$\gtrsim2.79\pm0.98$  &  $18.7$    & $111$  &   \\ 
090227B 	&  $1.61$  &  $5.89\pm0.30$   &  $28.3\pm1.5$     & $8921$ &  $<2.56$                  &  $71$      & $30$ \\ 
090510A  	&  $0.903$ &  $7.89\pm0.76$   &  $3.95\pm0.21$    & $9334$, $9336$ &$\gtrsim5.78\pm0.60$  &  $13.6$    & $1897$ &  \      & possible \textbf{kilonova} \\
140402A 	&  $5.52$  &  $6.1\pm1.6$     &  $4.7\pm1.1$      & $16069$, $16070$ &$\gtrsim4.5\pm2.2$    &  $13$      & $45$   &  \\ 
140619B 	&  $2.67$  &  $5.34\pm0.79$   &  $6.03\pm0.79$    & $16419$, $16420$ &$\gtrsim2.34\pm0.91$  &  $32$      & $149$  &   \\ 
160829A 	&  $4.373$  &  $0.92\pm 0.34$   &  $2.56 \pm 0.22$    & $19879$ &$\gtrsim 3.39\pm 2.95$  &  $14.8$      & $30.5$  &   \\ 
\hline
\end{tabular}
\caption{\textit{\textbf{List of 6 S-GRBs within the Fermi-LAT boresight angle. All of these events have associated  GeV photons}}: prompt and GeV emission properties of S-GRBs. In the first column we indicate the name of these sources. The second column gives their redshift. The third column gives the $E_{\rm p,i}$  deduced from the best fit model of the $Fermi$-GBM and remarkably  different from  those observed in S-GRFs. The fourth column gives the associated \textit{Fermi} GCN number. In {column 5} we estimate the characteristic $E_{\rm iso}$, which is systematically larger than the lower limit of $10^{52}$~erg. Column 6 shows the value of  $E_{\rm LAT}$ deduced from the $Fermi$-LAT. Column 7 shows the position of the source from the LAT boresight $\theta$. Column 8 is the likelihood test statistic (TS).}
\label{tab:bb}
\end{table*}

\begin{figure}
\centering
\includegraphics[width=1\hsize,clip]{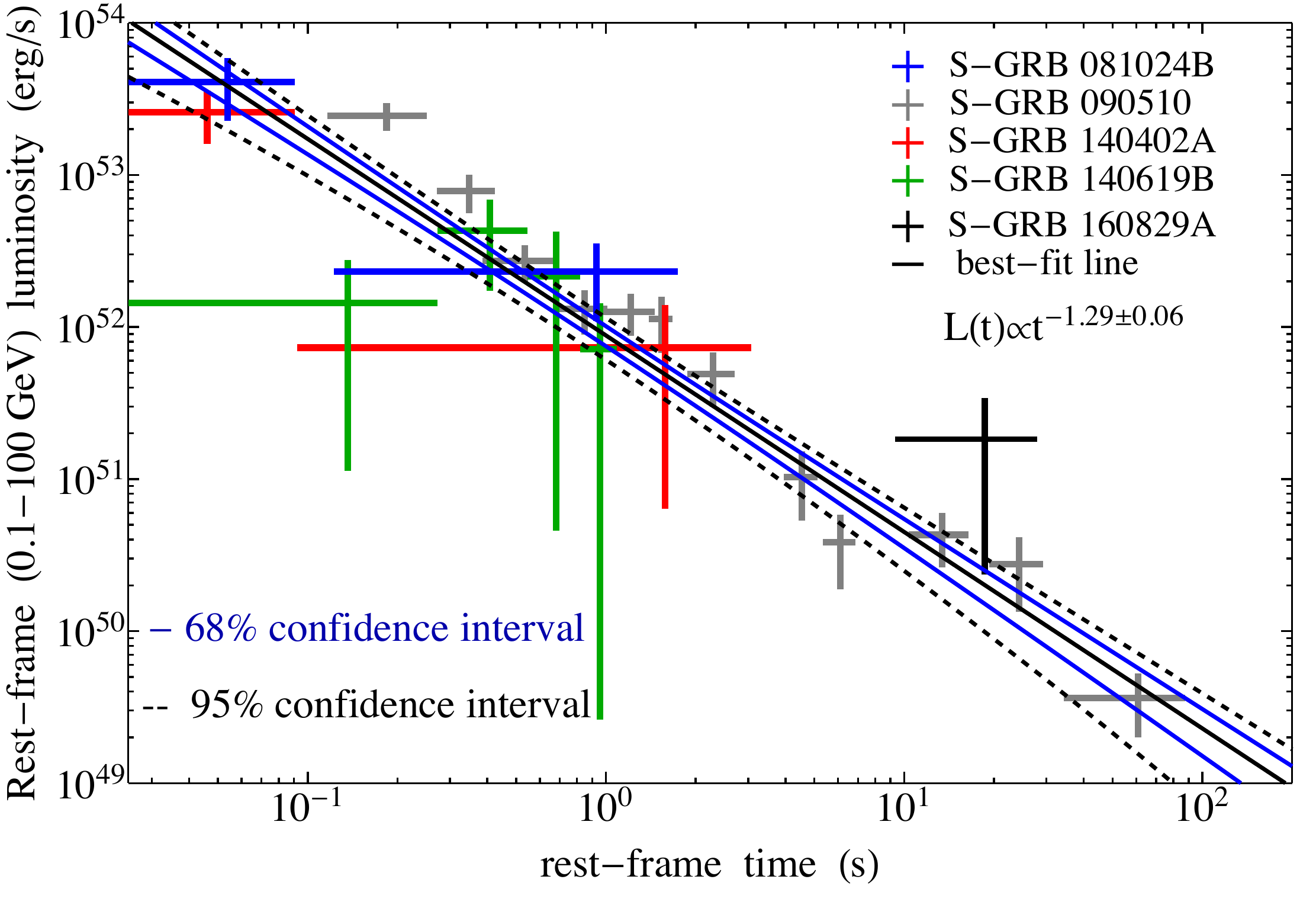}

\caption{The rest-frame $0.1$--$100$~GeV isotropic luminosity light-curves of all S-GRBs with LAT emission.  The black line indicate the common power-law behavior of the GeV emission with the slope of $\gamma=-1.29\pm0.06$.}
\label{fig:01}
\end{figure}

In previous papers \citep{2016ApJ...831..178R, 2017ApJ...844...83A} we introduced a specific luminosity dependence as a function of the rest-frame time of the GeV emission. This approach followed the corresponding one for the long GRBs originally introduced in \citet{2016ApJ...833..159P}. The rest-frame $0.1$--$100$~GeV isotropic luminosity light-curves follow a power-law behavior
\begin{equation}
L(t)= A~ t^{\gamma} 
\end{equation}
where $t$ is the rest-frame time  and $A = L(1s) $ is  the luminosity at 1 second. This has been well established in all S-GRBs, see, e.g. \citep{2014MNRAS.443.3578N,2015ApJ...798...10R,2015ApJ...808..190R} and here is {illustrated} in Fig.~\ref{fig:01}.

GRB 090510 is the S-GRB showing the longest GeV light-curve, lasting for $100$~s in its cosmological rest-frame. The other sources have some data points overlapping with GRB 090510.

By fitting all data points together using a common power-law, we confirm the  power-law index $\gamma=-1.29\pm 0.06$ and the luminosity at $1$~second $L(1 {\rm s}) = (0.88 \pm0.13) \times 10^{52}$~ergs~$s^{-1}$ (see Fig.~\ref{fig:01}), as previously obtained for {the} first 5 sources in \citet{2017ApJ...844...83A}.  We can see that, apart from the S-GRB 090227B, which was outside the nominal LAT field of view (FoV, i.e., $\theta\leq 65^\circ$), all S-GRBs exhibit GeV emission. This fact  points to the almost spherical symmetry or to a large  angle emission for the  GeV radiation  from all  these sources. 

In support of the classification of short bursts in two subclasses we confirm that the GeV emission is uniquely observed in S-GRBs and is related to the presence of the BH.

\section{The GeV emission and the energy extractable from a Kerr BH}\label{sec:5}

The first proposal for identifying the rotational energy  of the newly formed BH as {giving rise to} the GeV emission was indicated in the case of S-GRB 090510 \citep{2016ApJ...831..178R}. For this case, as well as for the S-GRBs 081024B and 140402A \citep{2017ApJ...844...83A}, we proposed that the GeV energy budget, $E_{\rm LAT}$, can be explained by the mass-accretion process onto the newly born Kerr BH. From this assumption we estimated lower limits on the amount of mass needed to explain $E_{\rm LAT}$ as due to the gravitational energy gained by accretion. Therefore, we adopted $E_{\rm LAT}= \eta_\pm M_{\rm acc}^{\eta_\pm} c^2$, where $M_{\rm acc}^{\eta_\pm}$ is the amount of accreted mass corresponding to the choice of the parameter $\eta_\pm$, that is the efficiency of the conversion of gravitational energy into GeV radiation. The value of $\eta_\pm$ depends on whether the infalling material is in a co- (+ sign) or counter-rotating (-- sign) orbit with respect to the BH and attains a maximum value of $\eta_+=0.42$ and $\eta_-=0.038$. This method successfully indicated {that} the accretion energy onto a Kerr BH is {sufficient} to explain the entire observed GeV emission. It was not, however, sufficient to infer either the BH mass or spin.

We now introduce an alternative procedure which  again is a sufficient condition to explain the GeV energetics, but also constrains both the mass and spin of the BH. Namely, we verify the condition that the GeV emission observed in the S-GRBs can originate from the BH extractable energy. We will address the same procedure for the BdHNe in {an} accompanying paper \citep{Ruffini2018arXiv180305476R}. In geometric units ($c=G=1$) and introducing the dimensionless intrinsic  angular momentum  $\alpha=J/M^2$ (the spin), the irreducible mass $M_{\rm irr}$ of a Kerr BH with mass $M$ and angular momentum $J$ can be obtained from the BH mass-formula \citep[see Eq.~2 in][]{1971PhRvD...4.3552C}:
\begin{equation}
\label{BHmass}
M^2=M_{\rm irr}^2+\alpha^2\frac{M^4}{4M_{\rm irr}^2}\ .
\end{equation}
Solving Eq.~(\ref{BHmass}) to find $M_{\rm irr}$, the extractable energy $E_{\rm extr}$, which is a function of $M$ and $\alpha$ given by 
\begin{equation}
\label{Eextr}
E_{\rm extr}=M-M_{\rm irr}=\left(1-\sqrt{\frac{1+\sqrt{1-\alpha^2}}{2}}\right)M\ .
\end{equation}
For an extreme Kerr BH ($\alpha=1$) we recover the well-known results that $E_{\rm extr}\approx 0.29 M$. For the maximum spin parameter attainable by a rotating NS, $\alpha_{\rm max}\approx0.7$,\footnote{This value is independent on the NS EOS; see e.g., \citet{2015PhRvD..92b3007C}.} the maximum extractable energy is smaller: $E_{\rm extr}\approx 0.074 M$. 

Assuming $E_{\rm LAT} = E_{\rm extr}$, we can obtain $M$ as a function of $\alpha$,  $M(\alpha)$. Since in S-GRBs the BH is formed out of a NS-NS merger, we request the condition $M(\alpha) \geq M_{\rm crit}(\alpha)$, where $M_{\rm crit}(\alpha)$ is the NS critical mass against gravitational collapse to a BH. It is easy to verify that this equality sets the minimum mass and, correspondingly, the maximum $\alpha$ for the newly-formed BH.

\begin{figure}
\centering
\includegraphics[width=\hsize,clip]{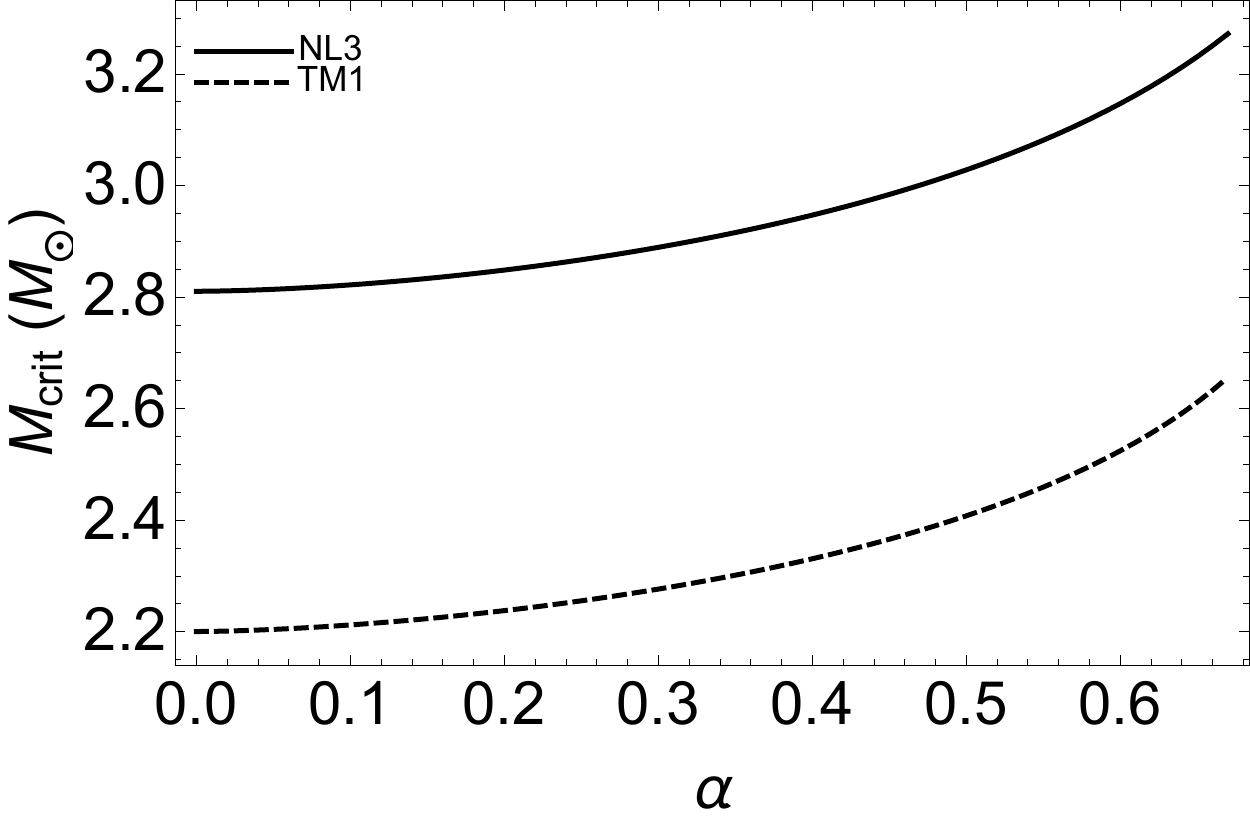}
\caption{NS critical mass as a function of the spin parameter $\alpha$ for the NL3 and TM1 EOS. We recall that the maximum spin parameter of a uniformly rotating NS is $\alpha_{\rm max}\approx0.7$, independently of the NS EOS; see e.g. \citet{2015PhRvD..92b3007C}.}\label{fig:Mcrit}
\end{figure}

The NS critical mass $M_{\rm crit}(\alpha)$ is given by the secular axisymmetric instability limit, when the star becomes unstable with respect to axially-symmetric perturbations. Following the turning-point method {of} \citet{1988ApJ...325..722F}, it is given by the critical point of a constant angular-momentum sequence of increasing central density, $\rho_c$, i.e. when $\partial M/\partial \rho_c = 0$. For instance, for the NL3, GM1 and TM1 EOS, \citet{2015PhRvD..92b3007C} obtained a formula that fits the numerical calculation results with a maximum error of 0.45\%, the numerical calculation results:
\begin{equation}\label{eq:Mcrit}
M_{\rm crit}=M_{\rm crit}^{J=0}(1 + k j^p),
\end{equation}
where $k$ and $p$ are parameters that depend upon the nuclear EOS, $M_{\rm crit}^{J=0}$ is the critical mass in the non-rotating (spinless) case and $j$ is the dimensionless angular momentum parameter $j \equiv \alpha (M_{\rm crit}/M_\odot)^2$.
For illustration we limit our investigation here, for the sake of example, to two nuclear EOSs: the TM1 EOS that leads to $k=0.017$, $p=1.61$ and $M_{\rm max}^{J=0}=2.20$~M$_\odot$, and to the very stiff NL3 EOS, that leads to $k=0.006$, $p=1.68$ and $M_{\rm crit}^{J=0}=2.81$~M$_\odot$ \citep{2015PhRvD..92b3007C}. The maximum value of $M_{\rm crit}$ is given by the secularly unstable configuration that is also maximally rotating, i.e. with $\alpha = \alpha_{\rm max}\approx 0.7$. The Eq.~(\ref{eq:Mcrit}) is a non-linear algebraic equation which cannot be solved analytically to obtain the  $M_{\rm crit}(\alpha)$ relation. Fig.~\ref{fig:Mcrit} shows the numerical solution of Eq.~(\ref{eq:Mcrit}) for the NL3 and TM1 EOS.

For the TM1 model, Eq.~(\ref{eq:Mcrit}) leads to a maximum critical-mass value $M_{\rm crit}(\alpha_{\rm max})=2.62$~M$_\odot$ (see Fig.~\ref{fig:Mcrit}). For this, Eq.~(\ref{Eextr}) gives a corresponding extractable energy $E_{\rm extr}=3.60\times10^{53}$~erg. For the NL3 model, we obtain $M_{\rm crit}(\alpha_{\rm max})=3.38$~M$_\odot$ (see Fig.~\ref{fig:Mcrit}) and $E_{\rm extr}=4.65\times10^{53}$~erg.

We can now apply the above procedure {to} the values of $E_{\rm LAT}$ in Table~\ref{tab:bb} to constrain the mass and spin of the newborn BH in S-GRBs.

Table~\ref{tab:cr} lists the inferred values of the minimum mass and corresponding maximum spin of the {BH inferred} from this procedure for both the TM1 and the NL3 nuclear EOS. The large uncertainty {in} the LAT energy mainly affects the estimate of the dimensionless angular momentum and also, but not so sensitively, on the estimate of the mass. This can be easily noted from Fig.~\ref{fig:Mcrit}: a small ranges of mass {translates into} large intervals of the $\alpha$ axis, at least for $\alpha<0.4$ which is the case for all S-GRBs.
 
\begin{table}
\centering
\begin{tabular}{lcccc}
\hline\hline
                &  \multicolumn{2}{c}{TM1}                                             &  \multicolumn{2}{c}{NL3} \\
\cline{2-5}
Source          &  $\alpha$                        &   $M(\alpha)$                     &  $\alpha$                           &   $M(\alpha)$ \\
        	    &                                  &   (M$_\odot$)                     &                                     &   (M$_\odot$) \\
\hline
S-GRB 081024B   &  $0.23^{+0.04}_{-0.04}$          &  $2.25^{+0.01}_{-0.01}$           &  $0.21^{+0.03}_{-0.04}$             &  $2.85^{+0.01}_{-0.01}$ \\
S-GRB 090227B 	&   --                             &  --                               &  --                                 &  --  \\
S-GRB 090510A  	&  $0.33^{+0.02}_{-0.02}$          &  $2.29^{+0.01}_{-0.01}$           &  $0.30^{+0.01}_{-0.01}$             &  $2.89^{+0.01}_{-0.01}$ \\
S-GRB 140402A 	&  $0.29^{+0.06}_{-0.08}$          &  $2.27^{+0.03}_{-0.03}$           &  $0.26^{+0.05}_{-0.07}$             &  $2.87^{+0.03}_{-0.03}$ \\
S-GRB 140619B 	&  $0.21^{+0.04}_{-0.05}$          &  $2.24^{+0.01}_{-0.02}$           &  $0.19^{+0.04}_{-0.05}$             &  $2.85^{+0.01}_{-0.01}$ \\
S-GRB 160829A   &  $0.29^{+0.10}_{-0.18}$          &  $2.27^{+0.05}_{-0.06}$           &  $0.26^{+0.09}_{-0.17}$             &  $2.87^{+0.04}_{-0.05}$ \\
\hline

\end{tabular}
\caption{Minimum mass and corresponding maximum spin parameters of the newly-formed {BH inferred} from the values of $E_{\rm LAT}$ for all S-GRBs in Fig.~\ref{fig:01}, and for the TM1 and the NL3 nuclear equations of state.}
\label{tab:cr}
\end{table}

We can then conclude that the formation of a Kerr BH is sufficient to explain the observed GeV emission, either by {an} accretion process or by using the BH extractable energy. This has been shown to {occur only} in S-GRBs.

\section{conclusion}\label{sec:6}

In this article we have discussed {whether} the existence of the GeV radiation can be used as a further confirmation and an additional means to discriminate between the S-GRF and S-GRB subclasses, along with their different energetics and spectra (S-GRF with $E_{\rm iso}\lesssim$ $10^{52}$~erg and $E_{\rm p,i}\sim0.2$--$2$~MeV; S-GRBs with $E_{\rm iso}\gtrsim 10^{52}$~erg, $E_{\rm p,i}\sim 2$--$8$~MeV). The GeV emission gives the necessary condition for the formation of the BH in the merging of the two NSs. In turn, the presence of the BH is sufficient to explain the entire energetics of the GeV emission by the BH mass and spin. The new feature identified in this article is the existence of the universal power-law behavior of S-GRBs GeV emission luminosity as a function of time in the rest-frame of the source, with amplitude $(0.88\pm 0.13) \times 10^{52}$ and a power-law index $1.29 \pm 0.06$. Such a relation makes it possible, in principle, to follow quantitatively the energy {production} process in S-GRBs and their temporal evolution, as well as identifying the essential parameters of the BH, namely its mass and intrinsic spin.

The great interest {in} short bursts has been {enhanced by the refined observations} of kilonova. This points to the possible occurrence of a kilonova within such  profoundly different binary systems as S-GRBs, S-GRFs and GRFs. This promises to be a  fertile topic for further investigation \citep{2018.Rueda...}. 

In an accompanying paper \citep{Ruffini2018arXiv180305476R} we further examine the role of GeV emission for long-duration GRBs. By examining $48$ XRFs and $329$ BdHNe, we show how the GeV emission provides an additional criteria for {distinguishing} between these two subclasses. The universal power-law behavior of $0.1$--$100$~GeV luminosity light-curves as a function of time   is confirmed for a subset of BdHNe. In particular, by using the mass formula of the Kerr BH, for BdHNe we can infer a minimum BH mass in the  $2.21$--$29.7 M_\odot$ range and a corresponding maximum dimensionless spin in the range of $0.07$--$0.71$. Most importantly, using the GeV data from \textit{Fermi}-LAT we can infer a new morphology of the SN ejecta accreting onto the newly-formed BH within the binary system.

Finally, it is also appropriate to mention that the results obtained here have many implications regarding the possible detection of gravitational waves coming from S-GRBs and S-GRF, especially concerning the potential connection with kilonovae emission, as recalled in section~\ref{sec:4}.
These are the subject of two forthcoming papers (\citet{2018.Rueda...} and {the} accompanying paper \citet{Ruffini2018arXiv180305476R}).

\acknowledgments
This work made use of data supplied by the UK Swift Science Data Center at the University of Leicester. Y.~A. is supported by the Erasmus Mundus Joint Doctorate Program Grant N.~2014-0707 from EACEA of the European Commission.
Y.~A. acknowledges partial support by the Targeted Financing Program BR05336383 of Aerospace Committee of the Ministry of Defense and Aerospace Industry of the Republic of Kazakhstan.
{M.~M. acknowledges partial support provided by a targeted funding for scientific and technical program of the Ministry of Education and Science of Kazakhstan.}
G.~J.~M is supported by the U.S. Department of Energy under Nuclear Theory Grant DE-FG02-95-ER40934.

\end{document}